\documentclass[preprint,showpacs]{revtex4}
\usepackage{graphics}
\usepackage{epsfig}
\usepackage{graphicx}
\usepackage{epstopdf}
\usepackage{bm}

\usepackage{color}

\begin{document}
\title{High-fidelity quantum driving}
\author{Mark G. Bason$^1$, Matthieu Viteau$^1$, Nicola Malossi$^2$, Paul Huillery$^{1,3}$, Ennio Arimondo$^{1,2,4}$, Donatella Ciampini$^{1,2,4}$,
        	     Rosario Fazio$^5$, Vittorio Giovannetti$^5$,
        	     Riccardo Mannella$^4$ \& Oliver Morsch$^1$}
\affiliation{$^1$INO-CNR, Largo Pontecorvo 3, 56127 Pisa, Italy\\ $^2$CNISM UdR, Dipartimento di Fisica `E. Fermi', Universit\`a di
Pisa, Largo Pontecorvo 3, 56127 Pisa, Italy\\$^3$ Laboratoire Aim\'{e} Cotton, Univ Paris-Sud 11, Campus d'Orsay Bat. 505, 91405 Orsay, France \\$^4$Dipartimento di Fisica `E. Fermi', Universit\`a di
Pisa, Largo Pontecorvo 3, 56127 Pisa, Italy\\ $^5$NEST,  Scuola Normale Superiore, and Istituto di Nanoscienze - CNR, 56126 Pisa, Italy}

\begin{abstract}
The ability to accurately control a quantum system is a fundamental requirement in many areas of modern science~\cite{wamsley} such as quantum information processing~\cite{nielsen} and the coherent manipulation of molecular systems. It is usually necessary to realize these quantum manipulations in the shortest possible time in order to minimize decoherence, and with a large stability against fluctuations of the control parameters.  While optimizing a protocol for speed leads to a natural lower bound in the form of the quantum speed limit rooted in the Heisenberg uncertainty principle~\cite{peres,levitin,bhattacharyya,caneva}, stability against parameter variations typically requires adiabatic following of the system. The ultimate goal in quantum control is to prepare a desired state with $100\%$ fidelity. Here we experimentally implement optimal control schemes that achieve nearly perfect fidelity for a two-level quantum system realized with Bose-Einstein condensates in optical lattices~\cite{Morsch2006}. By suitably tailoring the time-dependence of the system's parameters, we transform an initial quantum state into a desired final state through a short-cut protocol reaching the maximum speed compatible with the laws of quantum mechanics. In the opposite limit
we implement the recently proposed transitionless superadiabatic protocols~\cite{Lim1991,Demirplak2003,Demirplak2008,Berry2009}, in which the system perfectly follows the instantaneous adiabatic ground state. We demonstrate that superadiabatic protocols  are extremely robust against parameter variations, making them useful for practical applications
\end{abstract}

\maketitle

The need to control the dynamics of a quantum system is common to many different areas of science, ranging from the coherent
manipulation of molecular systems~\cite{molecular} to high precision measurements~\cite{precision}. Quantum control
may aim at reaching a given target state, as in the cooling of atomic ensembles~\cite{coolingatoms} and nano-mechanical
oscillators~\cite{coolingnems}, or at tracking the instantaneous ground state of a system during its evolution as in adiabatic quantum computation~\cite{farhi}. For each of these tasks an optimum strategy can be designed. In most applications the requirements on the fidelity of the final state is stringent, and in principle one aims at perfect fidelity.

Here we investigate high-fidelity quantum control protocols for the `simplest non-simple quantum problem'~\cite{berry_95a}, i.e. the evolution of a two-level system in a time $T$ as illustrated in Figure 1 (energies and times are expressed in the natural units of energy and time of our physical system, see Methods). The two diabatic levels $|0\rangle$ and $|1\rangle$ with energies $\Gamma(\tau)=\pm4(\tau-\frac{1}{2})$, where $\tau=t/T$ ranges from $0$ to $1$, are coupled through a coupling parameter $\omega$ giving rise to the adiabatic levels $|\psi_{g,e}(\tau)\rangle$. While there are infinitely many paths in Hilbert space connecting an initial quantum state $|\Psi_{ini}\rangle$ (for $\tau=0$) with a final target state $|\Psi_{fin}\rangle$ (for $\tau=1$) such that the final fidelity ${\cal F}_\mathrm{fin}=|\langle\Psi_{fin}|\psi_{g}(\tau=1)\rangle|^2=1$, in this work we concentrate on  two special classes. We consider those paths that minimize the time $T$ of the transformation and hence reach the quantum speed limit, and those paths ensuring a perfect following of the adiabatic ground state $|\psi_{g}(\tau)\rangle$. We will call the latter `superadiabatic' paths with reference to Berry's work on transition histories and the superadiabatic basis~\cite{Lim1991}.

We realize an effective two-level system using Bose-Einstein condensates (BECs) in an accelerated optical lattice (see Methods). Under appropriate conditions the wavefunction of the BEC in the periodic potential of the optical lattices can be approximated by considering only the two lowest energy bands~\cite{Zenesini2009}. Inside the first Brillouin zone the time-dependent Hamiltonian of the system can then be written in terms of the Pauli matrices $\sigma_i$ as
\begin{equation}
	{\cal H} =  \Gamma (\tau) \sigma_z+ \omega(\tau) \sigma_x \;.
\label{Hamiltonian}
\end{equation}
The system is initially prepared in the state $|\Psi_{ini}\rangle$, and the target is to reach $|\Psi_\mathrm{fin}\rangle$ after an evolution of duration $T$. The initial and final values of the parameters are chosen to be on opposite sides of the energy anticrossing (at $\tau=0.5$) (see Fig. 1).

We begin by considering a protocol that takes the system from $|\Psi_{ini}\rangle$ to $|\Psi_\mathrm{fin}\rangle$ with $100\%$ fidelity in the shortest possible time $T_{min}$. By analogy with the equivalent classical case this kind of protocol has been called the `quantum brachistochrone'~\cite{carlini,caneva}. If we only impose the constraint that $\omega$ is constant (otherwise $T_{min}\rightarrow 0$ as $\omega\rightarrow\infty$), we find that the protocol shown in Figure 2c minimizes $T$. This `composite pulse' protocol (see Methods), in close analogy to composite pulses in NMR~\cite{levitt1996}, represents half a Rabi oscillation with frequency $\omega$ at $\tau=0.5$, preceded and followed by two short pulses (in theory delta-functions) with a pulse area of $\pi/4$ (see also~\cite{mellish2003}). The experimentally measured minimum times (Fig. 2d) for reaching the target state with fidelity ${\cal F}_{fin}\approx1$ approach the quantum speed limit, also known as the Fleming or Bhattacharyya bound, given by
\begin{equation} \label{qsoptimal}
T_{\rm qs}=\frac{\arccos|\langle\Psi_{fin}|\Psi_{ini}\rangle|}{\omega}.
\end{equation}

One can assess the performance of the composite pulse protocol by comparing it to the paradigmatic Landau-Zener (LZ) protocol and to the locally adiabatic protocol proposed by Roland and Cerf in the context of adiabatic quantum computation~\cite{roland}. In the LZ problem with constant $\omega$ and linearly varying $\Gamma(\tau)$ a system initially prepared in the adiabatic ground state undergoes tunneling to the excited state with a finite probability, leading to a fidelity ${\cal F}_{fin}=\exp(-\frac{\pi T\omega^2}{4})<1$. Stricter boundary conditions are imposed in the approach taken by Roland and Cerf.  On top of the condition that $\omega$ be constant they demand local adiabatic following to within some small deviation $\epsilon$, i.e. during the entire protocol ${\cal F}(\tau)\geq1-\epsilon^2$ (see Methods). In both the LZ and the Roland and Cerf protocols the time to achieve perfect adiabaticity, i.e. ${\cal F}_{fin}=1$, diverges and hence we measure the time to reach ${\cal F}_{fin}=0.9$ as a function of $\omega$ instead (shown as a dotted line in Fig. 2e). The results of the comparison are shown in Fig. 2d. It is evident that while the LZ protocol is more than an order of magnitude slower than the composite pulse protocol for small $\omega$, the Roland-Cerf protocol reaches ${\cal F}_{fin}=0.9$ in a time that is only $10-50\%$ above the lower limit given by the Fleming-Bhattacharyya bound.

At the opposite extreme of the quantum control spectrum, rather than minimizing the total time one can maximize the adiabaticity during the protocol. While the composite pulse protocol described above was obtained through optimization given a constraint on $\omega$, it is possible to analytically calculate protocols that ensure ${\cal F}(\tau)=1$ during the entire evolution. The reasoning behind such a transitionless super-adiabatic (or counter-adiabatic \cite{Demirplak2003,Demirplak2008}) protocol is that for a given time-varying Hamiltonian it is always possible to construct an auxiliary Hamiltonian ${\cal H}_s$ that cancels the non-adiabatic part of the evolution under ${\cal H}$ alone. It thus ensures transitionless adiabatic following such that the system evolving under ${\cal H}+{\cal H}_s$ always remains in the instantaneous adiabatic ground state of ${\cal H}$ with $100\%$ probability~\cite{Lim1991,Demirplak2003} for a finite duration of the protocol. In general ${\cal H}_s$ can be written as
\begin{equation}
{\cal H}_s(t)=i\hbar\sum_n|\partial_t n(t)\rangle\langle n(t)|,
\end{equation}
where $|n(t)\rangle$ are the eigenstates of the original Hamiltonian ${\cal H}$.
For a two-level system of the form (1) one finds that
\begin{equation}
{\cal H}_s(t)=\frac{\hbar}{2}\frac{\partial \phi}{\partial t}\sigma_y,
\end{equation}
where $\phi=\arctan \frac{\omega(t)}{\Gamma(t)}$.
This means that in order to make a given evolution of a two-level system perfectly adiabatic one needs to add an interaction term corresponding to a $\sigma_y$ Pauli matrix~\cite{Chen2010}. In practice, ${\cal H}_s$ can be implemented by introducing an additional interaction into the system, e.g. through an extra laser or microwave field. In the case of atoms in an optical lattice considered in this work, the additional Hamiltonian can be realized by adding a second optical lattice shifted with respect to the first one by $d_L/4$ ($d_L$ is the lattice spacing). It can be shown, however, that the effect of this additional field can also be achieved through an appropriate transformation $\Gamma\rightarrow\Gamma'$ and $\omega\rightarrow\omega'$ (see Methods), so that no extra field is necessary. This result, which is independent of the physical system under consideration, means that the resulting protocol is intrinsically more stable as there will be no problems associated, e.g, with phase fluctuations between the fields~\cite{Demirplak2003,Demirplak2008}.

For the standard LZ protocol with fixed $\omega$ and linearly varying $\Gamma(\tau)$ the general shape of the required transformation is shown in Figure 3a (for details see Methods), together with the result of an experiment in which for different sweep durations $T$ the fidelity ${\cal F}_{fin}$ was measured both for the linear LZ protocol and for the superadiabatic protocol. We find that in the superadiabatic case ${\cal F}_{fin}\gtrsim0.98$ for all $T$. Furthermore, a time-resolved measurement of ${\cal F}(\tau )$ during the sweep (Fig. 3d) shows that the system stays in the ground adiabatic state at all times.

In principle, the superadiabatic transformation of $\Gamma$ and $\omega$ can be calculated for any sweep protocol. There are, however, special cases that are of particular interest. For example, a protocol for which the correction in $\Gamma$ vanishes, i.e. for which $\Gamma'=\Gamma$ (except at the beginning and the end of the sweep, see Methods), can be expected to be more robust to variations in its parameters. This is the case for the superadiabatic `tangent protocol' (see Methods) shown schematically in Figure 3b. As in the case of the linear LZ protocol with superadiabatic corrections, in the tangent protocol the systems remains in the adiabatic ground state to within $1\%$ throughout the entire protocol. Since the noise in our imaging system does not allow us reliably to measure fidelities ${\cal F}_{fin}\gtrsim0.98$, we performed an experiment in which the tangent sweep was repeated four times, giving an overall probability ${\cal F}_{fin}^{tot}\approx0.94$, from which we deduce that, indeed, in a single sweep ${\cal F}_{fin}\approx0.99$. This figure is compatible with a fidelity of $100\%$ as we estimate the non-adiabaticity of the preparation and measurement protocols to be on the order of $1\%$.

In order to test the sensitivity of the superadiabatic protocols to a (simulated) variation in the control parameters, we varied both $\omega$ and $T$ around the optimum value and measured the fidelity ${\cal F}_{fin}$ in each case. The results are summarized in Figure 4a, which shows clearly that the tangent sweep is extremely robust with respect to an increase in $T$ or $\omega$, with ${\cal F}_{fin}\gtrsim0.99$ for increases up to $100\%$.

Finally, we compare the speed of the superadiabatic tangent protocol with the composite pulse protocol as a function of $\omega'$ (for the composite pulse protocol $\omega'=\omega$ being constant). Solving Eqn. 10 for $T$ gives a total time for the superadiabatic tangent protocol that depends on both $\omega$ and $\omega'$ (with $\omega'>\omega$). It is, therefore, possible to minimize $T$ for a given value of $\omega'$ by choosing an appropriate value for $\omega$. The result of this minimization is shown in Fig. 4b. Surprisingly, the minimum value of $T$ as a function of $\omega'$ for the superadiabatic tangent protocol lies below the LZ and Roland and Cerf times and is quite close to the quantum speed limit, meaning that the 'penalty' in terms of speed for the requirement of perfect adiabatic following is less than one might at first expect. In fact, for small values of $\omega'$ one can formally let $\omega\rightarrow0$ in the expression for $\omega'$ (see Eqn. 21 in Methods), giving $T=\frac{\pi}{2\omega'}$ which coincides with the quantum speed limit for orthogonal states with $|\langle\Psi_{fin}|\Psi_{ini}\rangle|=0$.

In summary, we have explored high-fidelity quantum control protocols for the evolution of an artificial two-level quantum system ranging from the speed-limited to the superadiabatic regime. The superadiabatic transformations make it possible to readily implement protocols ensuring perfect adiabatic following in a variety of existing applications. In practice, of course the choice of protocol will depend on the boundary conditions and physical limitations of the system under consideration. If both $\Gamma$ and $\omega$ can be controlled (to within some limits), the superadiabatic protocols provide the possibility of state preparation with $100\%$ fidelity with high stability against parameter variations.

\section*{Methods}
\subsection{Bose-Einstein condensates in optical lattices}

We realize an effective two-level system by loading Bose-Einstein condensates into optical lattice potentials \cite{Morsch2006} of the form $\frac{V_0}{2} \cos\left(2\pi x/d_L+\phi(t)\right)$. In the limit of small lattice depths $V_0\lesssim 5\,E_\mathrm{rec}$ (here the recoil energy $E_{\rm rec}=\hbar\omega_{\rm rec}=\pi^{2}\hbar^{2}/2 M d_{\rm L}^{2}$, and $\omega_{\rm rec}=2\pi\times 3.15\,\mathrm{kHz}$ defines the natural units of energy $\hbar\omega_{rec}$ and time $1/\omega_{rec}$ for our system), the lowest energy levels are given by the quadratic dispersion relations of free particles with momenta differing by $2\hbar k_L$ coupled through a coupling constant $\omega=V_0/4$. Subtracting the quadratic term in the momentum \cite{tayebirad2010} leads to the effective Hamiltonian (1), where the time dependence in $\tau$ is now achieved through a variation of the quasimomentum $q$ in the first Brillouin zone. Experimentally, $q$ and hence $\Gamma=4\hbar\omega_{rec}(q-\frac{1}{2})$ (in this Methods section we shall use the explicit physical units wherever appropriate) is controlled through the term $\phi(t)$ which can be used to accelerate the lattice, leading to an inertial force in the rest frame of the lattice. The time dependence of $\omega$ is controlled through the power of the lattice laser beams which determine $V_0$.

The experimental protocols are carried out using techniques previously developed by us and described in detail in~\cite{Zenesini2009,tayebirad2010}. Time-resolved measurements
with superadiabatic driving are performed by applying appropriate jumps of the lattice position, of the quasimomentum and/or of the lattice depth,  before measuring in the adiabatic basis (now of the original Hamiltonian $\cal{H}$).

\subsection{Driving protocols}
\subsubsection{Composite pulse protocol}

The optimality of the time~(\ref{qsoptimal}) in the LZ setting was derived in Ref.~\cite{caneva}
by starting from a guess function similar to the Roland and Cerf protocol (see next section), and by running a numerical search aimed at determining the minimal value of $T$ (in that optimization the adiabatic requirement was not enforced).
The optimal pulse $\Gamma(\tau)$ associated with such a minimal time corresponds to highly irregular functions which are strongly peaked at the beginning and at the end of the protocol while remaining flat and equal to zero for intermediate times.
Assuming that no constraints are imposed on  $\Gamma(\tau)$ and on its first derivative
 one can extrapolate the  exact (asymptotic) analytical form of such pulses as
 \begin{eqnarray} \label{ggg}
\Gamma(\tau) = \left\{
\begin{array}{ll}
-\Gamma_0   & \mbox{for $\tau=0$} \\
\Gamma_M  & \mbox{for $\tau\in[0,\tau_0]$} \\
0 & \mbox{for $\tau\in[\tau_0,1-\tau_0]$} \\
-\Gamma_M   & \mbox{for $\tau\in[1-\tau_0,1]$} \\
+\Gamma_0   & \mbox{for $\tau=1$}\;, \\
\end{array}\right.
\end{eqnarray}
where $\Gamma_0=-2$ as usual, while $\Gamma_M$ and $\tau_0$ are, respectively,   asymptotically large and small quantities which satisfy the condition
\begin{eqnarray}\label{oo}
 \Gamma_M \tau_0 = \pi/4\;.
\end{eqnarray}
For $\Gamma_M \gg 1$
the evolution described by the pulse~(\ref{ggg})
corresponds to first applying a fast (instantaneous) clock-wise $\sigma_z$ pulse around the $z$-axis to the system which
take the point on the Bloch sphere lying in the $x-z$ plane and, in the short time $\tau_0$, rotates it into the $y-z$ plane.
The system then undergoes a rotation at frequency $\omega$ around the $x-$axis for a time $1-2\tau_0$, while finally another instantaneous $\sigma_z$ pulse rotates the $x-z$ plane back
to its initial position.

The {\em exact} transfer of $|\Psi_{ini}\rangle$ to $|\Psi_{fin}\rangle$ is achieved by choosing $T$ such that
\begin{eqnarray}
T(1-2\tau_0) =  \frac{\arccos{|\langle \Psi_{fin}|\Psi_{ini}\rangle|}}{ \omega}\;,
\end{eqnarray}
which coincides
with the r.h.s.  of Eq.~(\ref{qsoptimal})
by letting $\tau_0\rightarrow 0$ while keeping~(\ref{oo}).
We note here that while the fact that the composite pulse protocol realizes that quantum speed limit suggests that this time is optimal, a formal proof for this is still missing.

\subsubsection{Roland-Cerf protocol}

In the following we summarize the analysis of Roland and Cerf~\cite{roland} adapting it to the case of the Hamiltonian ${\cal H}$ of  Eq.~(1) under the assumption that the coupling $\omega$ is kept constant.  In this protocol the requirement is to keep the evolution adiabatic in each infinitesimal time interval. At any instant of the evolution the fidelity of the state with the instantaneous ground state is required to be ${\cal F}(\tau)=  | \langle \psi(\tau) | \psi_g{(\tau)} \rangle|^2=1 - \epsilon^2$, from which it follows that the total time of the protocol is
\begin{equation}
T(\epsilon) =  \frac{1}{\epsilon \omega} \; \frac{1}{\sqrt{4 + \omega^2}} \;,
\end{equation}
for a time-dependence of $\Gamma(t)$ of the form
\begin{equation}
\Gamma(\tau) = \frac{ 4 \epsilon \omega^2 T(\epsilon) (\tau - 1/2)}{ \sqrt{ 1 - 16 \epsilon^2 \omega^2 T(\epsilon)^2 (\tau - 1/2)^2} } \;.
\end{equation}
(as shown in Fig. 2b in the main text).

\subsubsection{Superadiabatic protocol}
In order to implement superadiabatic (transitionless) driving, we recast the Hamiltonian ${\cal H}(\tau)+{\cal H}_s(\tau)=\Gamma(\tau)\sigma_z+\omega(\tau)\sigma_x+\frac{\hbar}{2}\frac{\partial \phi}{\partial t}\sigma_y$ in the form ${\cal H}'=\Gamma'(\tau)\sigma_z+\omega'(\tau)\sigma_x$, eliminating the need for an extra potential that realizes the $\sigma_y$-term. The necessary transformations $\Gamma\rightarrow\Gamma'$ and $\omega\rightarrow\omega'$ can be derived by observing that for the physical system used in this paper, i.e. matter waves in a periodic potential, the Pauli matrices $\sigma_x$ and $\sigma_y$ are simply related by the spatial displacement operator $\hat{U}_\mathrm{d}(\delta x)$ with $\delta x = d_L/4$, where $d_L$ is the lattice constant. In the basis of the plane waves $\exp\left(\pm ikx \right)$ defining the two-state subspace $\{|0\rangle,|1\rangle\}$, this operator can be written as
\begin{equation}
\hat{U}_\mathrm{d}(d_L/4) = \left( \begin{array}{cc} e^{\frac{\pi}{4}i} & 0 \\ 0 & e^{-\frac{\pi}{4}i} \end{array} \right)
\end{equation}
and hence
\begin{equation}
\sigma_y = \hat{U}^{\dag}_\mathrm{d}(d_L/4)\sigma_x \hat{U}_\mathrm{d}(d_L/4).
\end{equation}
This means that the $\sigma_y$-term in the transitionless driving protocol can be realized by adding a second periodic potential shifted by $d_L/4$ and accelerated in the same way as the first one. Since the sum of two periodic potentials of the same periodicity is again a periodic potential with a modified phase and amplitude, we can write the combined potential
\begin{equation}
V_\mathrm{tot}(x)=V_0\cos^2\left(\frac{\pi x}{d_L}\right)+4\alpha\cos^2\left(\frac{\pi (x-\frac{d_L}{4})}{d_L}\right),
\end{equation}
as
\begin{equation}
V_\mathrm{tot}(x)= const.+2\sqrt{\left(\frac{V_0}{2}\right)^2+(2\alpha)^2} \cos^2\left(\frac{\pi (x-\frac{\beta d_L}{2\pi})}{d_L}\right),
\end{equation}
where $\alpha = \frac{\hbar}{2}\frac{\partial \phi}{\partial t}$ and $\beta=\arctan \frac{4\alpha}{V_0}$.
From this equation, the transformation rule for $\omega=V_0/4$ can be read off immediately, while the time derivative of the displacement term $\frac{\beta d_L}{2\pi}$
gives the correction to the quasimomentum
\begin{equation}
q'(t)=q(t)-\frac{\dot{\beta}}{8\omega_\mathrm{rec}},
\end{equation}
from which the transformation rule for $\Gamma$ is calculated using the relation $\Gamma=4\hbar\omega_{rec}(q-\frac{1}{2})$.
The complete transformation then reads (in rescaled units)
\begin{equation}
\Gamma'=\Gamma-\frac{1}{2}\frac{d}{dt}\left[\arctan\left(\frac{\Gamma\dot{\omega}-\omega\dot{\Gamma}}{2\omega(\Gamma^2+\omega^2)}\right)\right],
\end{equation}
\begin{equation}
\omega'=\omega\sqrt{1+\frac{(\Gamma\dot{\omega}-\omega\dot{\Gamma})^2}{4\omega^2(\Gamma^2+\omega^2)^2}}
\end{equation}
where the discontinuities in the expression in square brackets in $\Gamma'$ at the beginning and at the end of the protocol give rise to delta-functions, which can be realized in practice using large but finite corrections $\Delta\Gamma_M$ for a short duration $\Delta t$ such that
\begin{equation}
\Delta t\Delta\Gamma_M=\mp\frac{1}{2}\arctan\left(\frac{\Gamma\dot{\omega}-\omega\dot{\Gamma}}{2\omega(\Gamma^2+\omega^2)}\right),
\end{equation}
where the $-$ and $+$ signs refer to the corrections at the beginning and at the end of the protocol, respectively.

The two superadiabatic protocols considered in this paper are the superadiabatic linear protocol with
\begin{eqnarray}
\Gamma'(\tau)&=\Gamma(\tau)-\frac{4(\tau-\frac{1}{2})}{T^2\left[(\tau-\frac{1}{2})^2+\frac{1}{2}\omega^2\right]^2+1},\\
\omega'(\tau)&=\omega\sqrt{1+\frac{1}{T^2(8(\tau-\frac{1}{2})^2+\frac{1}{2}\omega^2)^2}}
\end{eqnarray}
and the superadiabatic tangent protocol for which
\begin{equation}
\Gamma'(\tau)=\Gamma(\tau)=\omega\tan\left(2\left(\tau-\frac{1}{2}\right)\arctan\left(\frac{2}{\omega}\right)\right)
\end{equation}
(apart from the delta-functions at the beginning and at the end of the protocol) and
\begin{equation}
\omega'=\omega \sqrt{1+\frac{\arctan(\frac{2}{\omega})^2}{(T\omega)^2}}.
\end{equation}
The latter protocol is found by demanding that $\Gamma'(\tau)=\Gamma(\tau)$ and solving the resulting differential equation.

\bibliographystyle{naturemag}

\section*{Acknowledgments}
This work was supported by CNISM through the Progetto Innesco 2007, the E.U. through grant No. 225187-NAMEQUAM and the collaboration between the University of Pisa and the University Paris Sud-11. R.F. and V.G. acknowledge support by MIUR through FIRB-IDEAS Project No. RBID08B3FM and by the E.U. through grants No. 234970-NANOCTM and No. 248629-SOLID. The authors thank D. Gu\'{e}ry-Odelin and M. Holthaus for fruitful discussions.

\section*{Author contributions}
M.G.B., M.V., N.M., P.H. and D.C. carried out the experiments; V.G. and R.F. developed the composite pulse and Roland and Cerf protocols; O.M. and R.M. developed the superadiabatic protocols and performed the numerical simulations; E.A., R.F. and O.M. wrote the paper. All authors discussed the results and commented on the manuscript.

\newpage
\section*{Figure captions}

\begin{figure}[htp]

\caption{Schematic of a two-level quantum system with crossing energy levels. The bare (diabatic) states $|0\rangle$ and $|1\rangle$ are coupled to give the adiabatic states $|\Psi_g\rangle$ and $|\Psi_e\rangle$. The adiabatic states have an energy gap of $2\omega$ where the diabatic levels cross (at $\tau=0.5$, $E=0$). The arrows indicate the two extreme optimized protocols discussed in this paper: the `short cut' composite pulse protocol reaching the quantum speed limit (blue line) and the superadiabatic protocol (red line) for which the system perfectly follows the instantaneous ground state.}

\end{figure}

\begin{figure}[htp]

\caption{Comparison between the Landau Zener (a), Roland and Cerf (b) and composite pulse protocols (c). (d) The time needed to achieve a fidelity of $100\%$ for the composite pulse protocol (red triangles) and the minimum time to achieve ${\cal F}_{fin}=0.9$ in the Roland-Cerf protocol (grey circles) and the LZ protocol (blue squares). (e) Fidelity of the final state as a function of the duration for the composite pulse protocol (red triangles), the Roland-Cerf protocol (grey circles) and the LZ protocol (blue squares). Note that ${\cal F}_{fin}=1$ is not reached in the composite pulse protocol because of an interaction-induced loss of coherence in the BEC, which also leads to fidelities that are slightly below the theoretical predictions for the other protocols. The dashed lines are theoretical predictions. All experimental data are for $\omega=0.5$.}

\end{figure}

\begin{figure}[htp]

\caption{Superadiabatic dynamics in a two-level system. (a,b) Original (dashed lines) and superadiabatic (solid lines) protocols for the linear (a) and tangent (b) cases. (c) Fidelity as a function of the duration of the protocol for the superadiabatic linear (grey circles) and LZ protocols (blue squares). For comparison, the results of a linear LZ sweep in which only $\omega$ is transformed (Eq. 8) is also shown (empty squares). The dashed lines are theoretical predictions. (d) Fidelity during the protocol for the superadiabatic linear (grey circles), superadiabatic tangent (red squares) and LZ protocols (blue squares). Clearly, for the superadiabatic protocols ${\cal F}(\tau)\approx 1$ during the entire protocol (to within the experimental error). All experimental data are for $\omega=0.55$.}

\end{figure}

\begin{figure}[htp]

\caption{Robustness and speed of the superadiabatic tangent protocol. (a) Fidelity of the superadiabatic tangent protocol as a function of the relative deviation of $T$ (red squares) and $\omega$ (empty squares) from their ideal values $\omega=0.5$, $T=5.9$. The solid and dashed lines are numerical simulations. Inset: Detail of the graph, highlighting the stability of the protocol for $\Delta T/T,\Delta\omega/\omega>0$. (b) Comparison of the duration of superadiabatic tangent protocols with the composite pulse (dashed red line), Roland and Cerf (dashed black line) and Landau Zener protocols (dashed blue line). The solid black line shows the minimum time for the superadiabatic tangent protocol as a function of $\omega'$.}

\end{figure}

\newpage
\section*{Figures}
\setcounter{figure}{0}
\begin{figure}[htp]
\includegraphics[width=12cm]{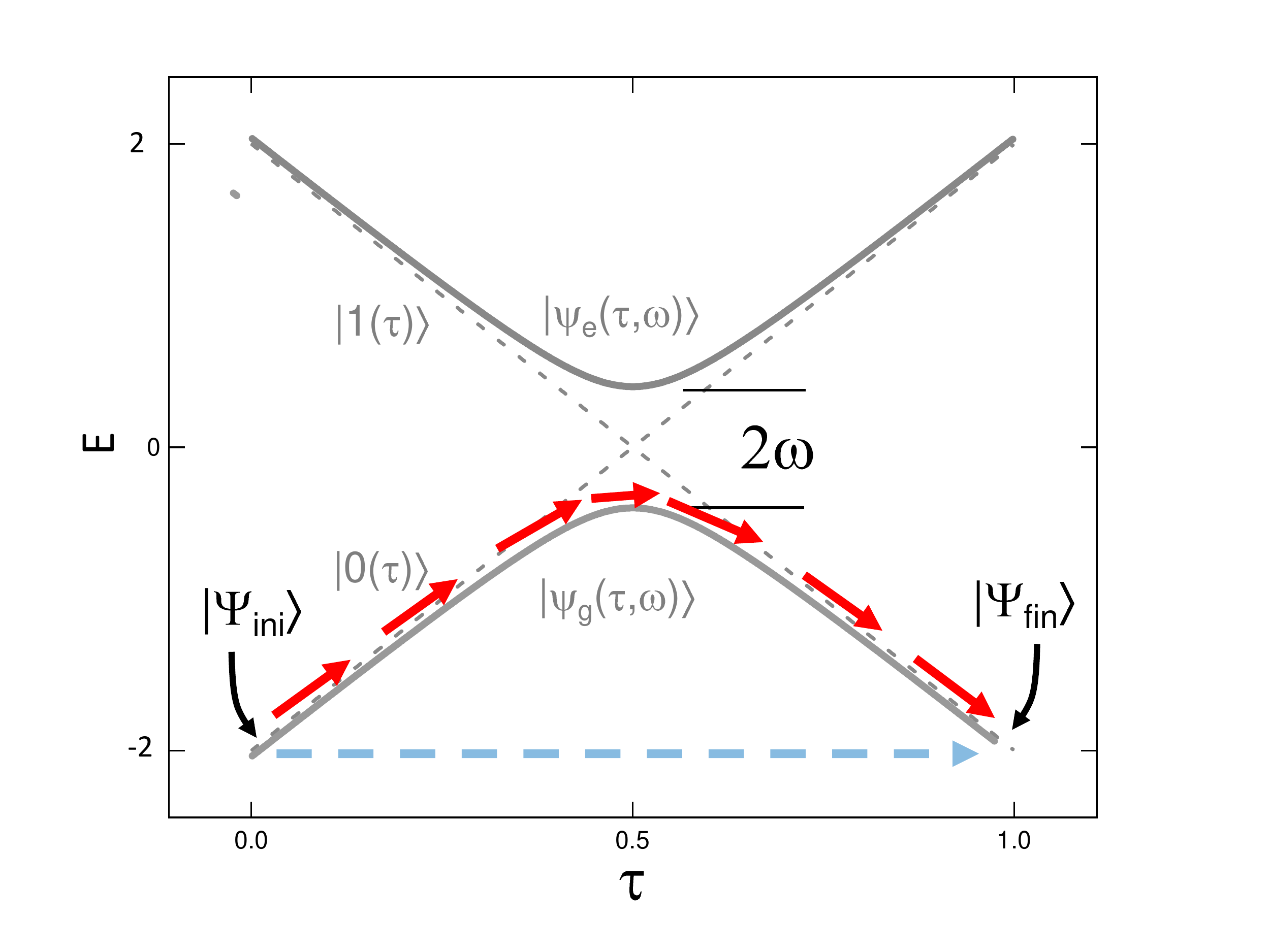}
\caption{}
\end{figure}
\newpage

\begin{figure}[htp]
\includegraphics[width=12cm]{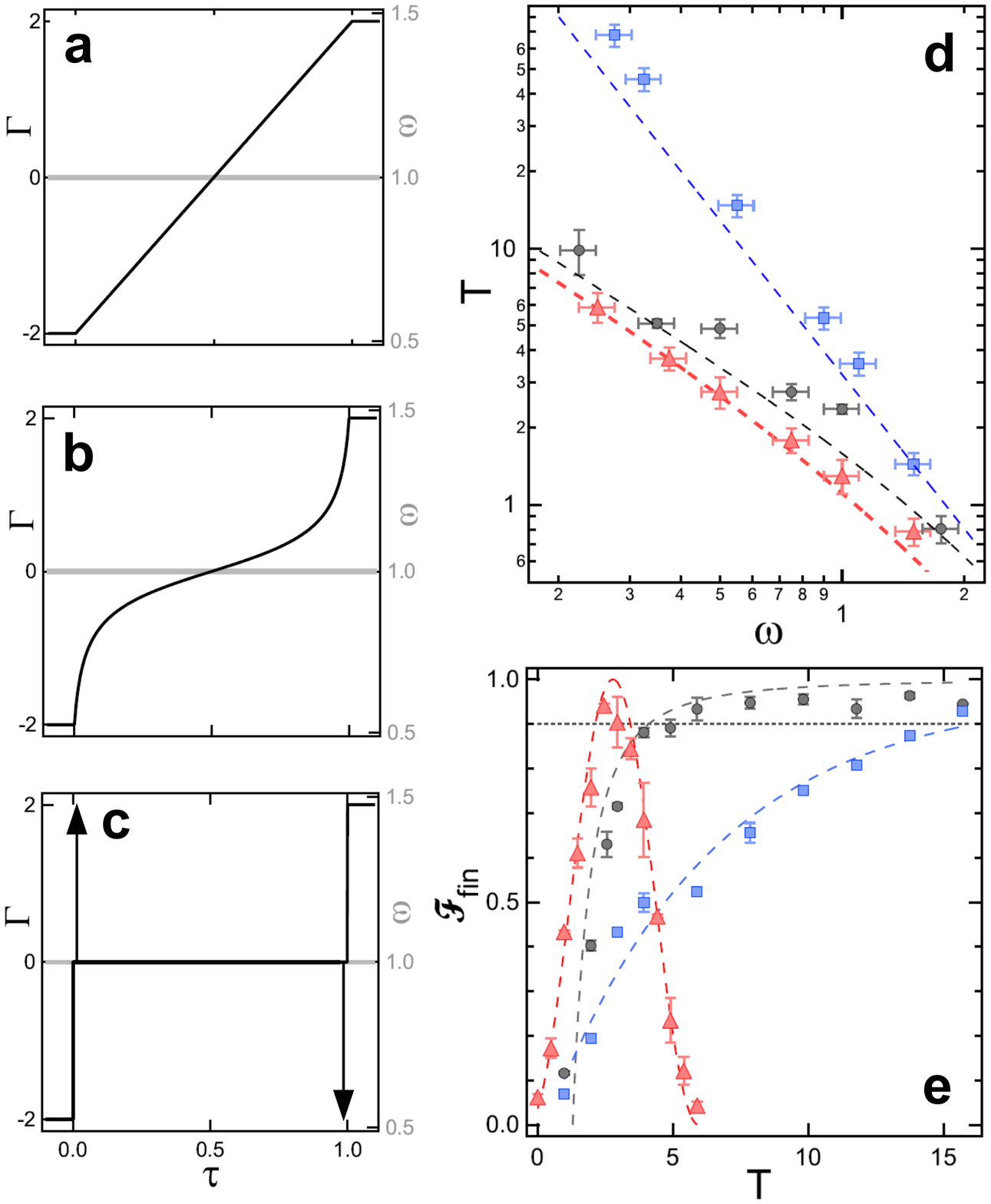}
\caption{}
\end{figure}
\newpage

\begin{figure}[htp]
\includegraphics[width=12cm]{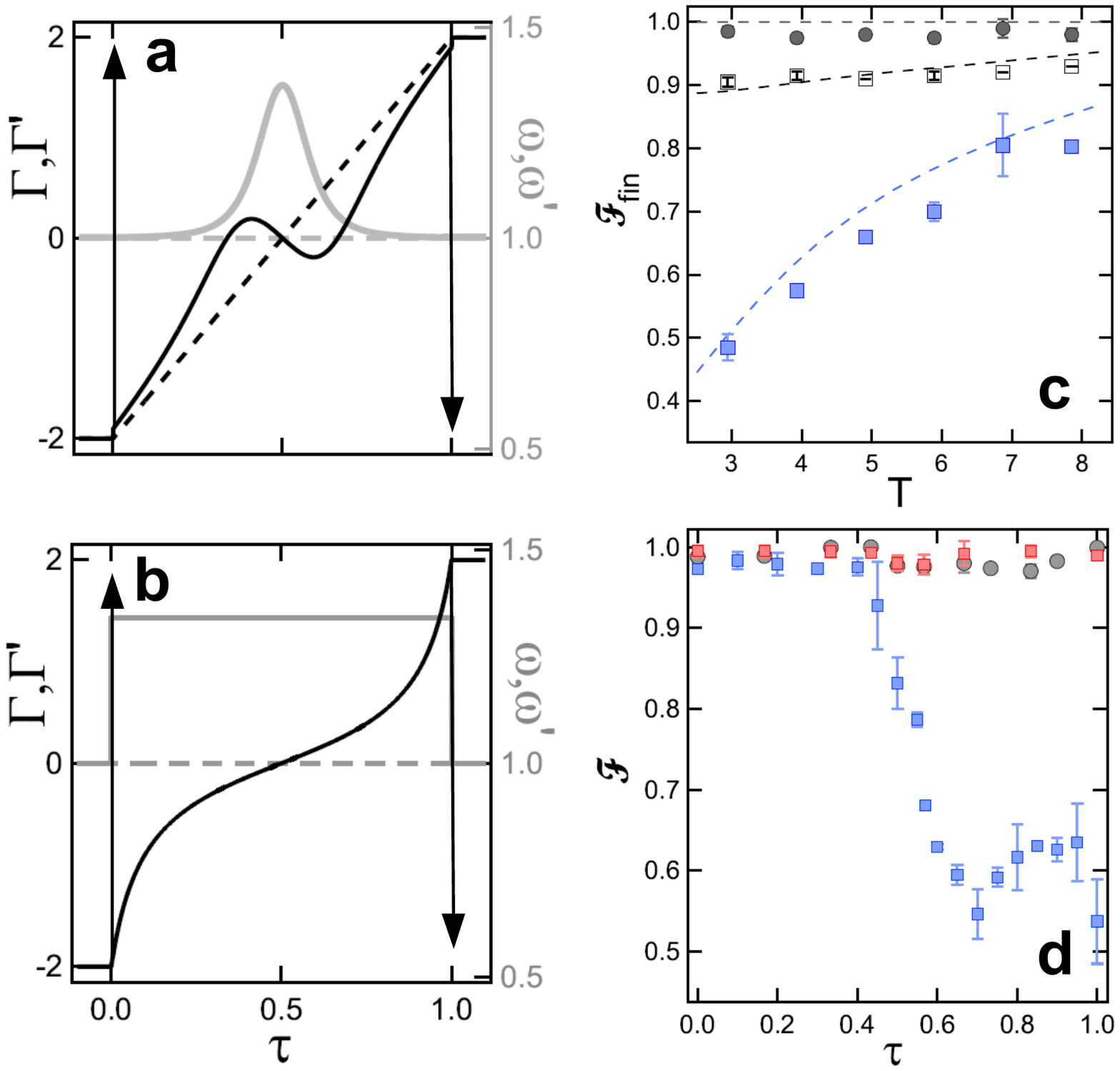}
\caption{}
\end{figure}
\newpage

\begin{figure}[htp]
\includegraphics[width=14cm]{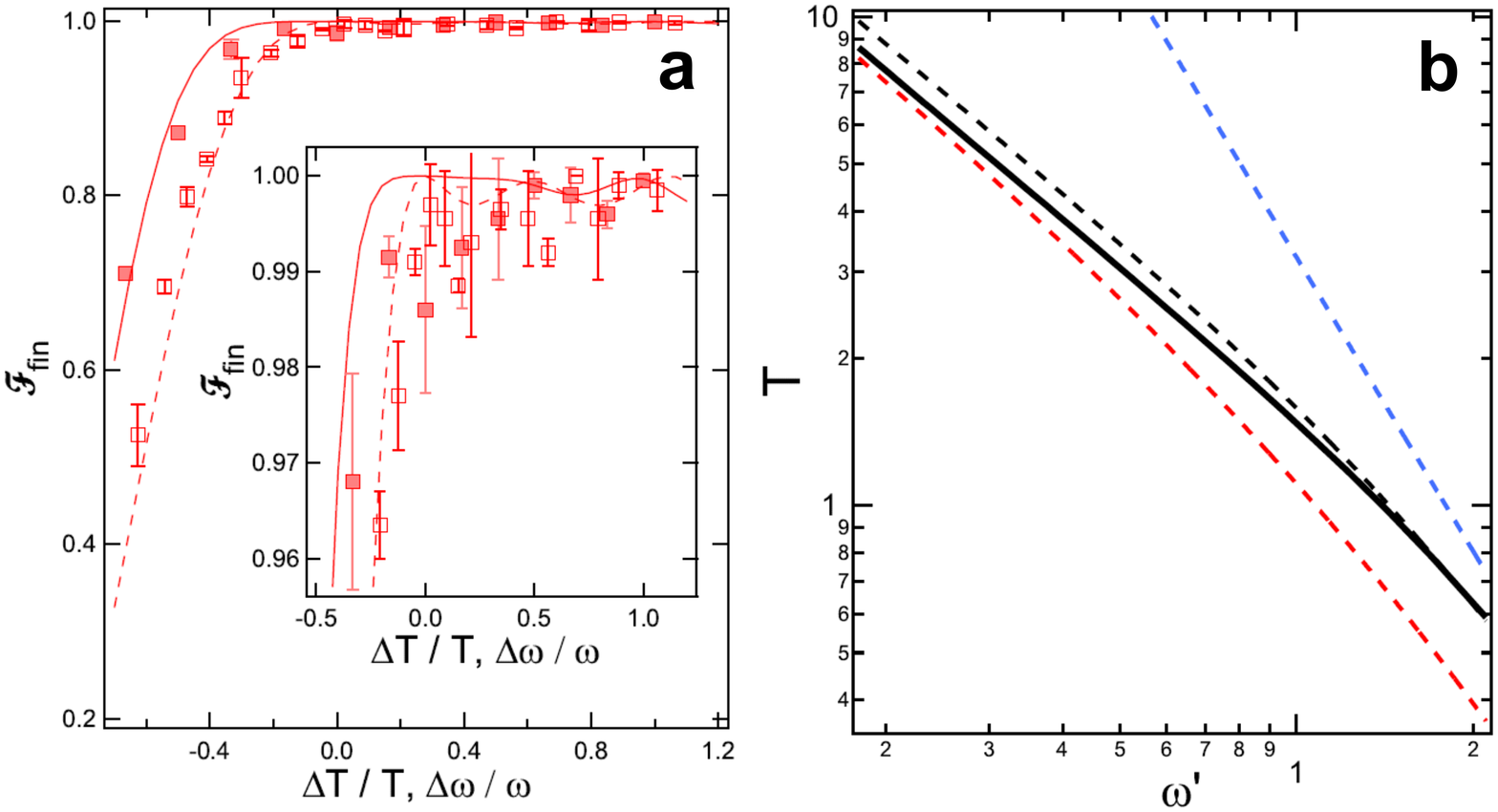}
\caption{}
\end{figure}

\end{document}